\documentclass{nature}

\usepackage{amsmath}
\usepackage[pdftex]{graphicx}
\usepackage{dcolumn}
\usepackage{bm}     
\usepackage{dsfont}
\usepackage{color}
\usepackage{braket}

\newcounter{defcounter}
\usepackage{amsthm}

\title{Photoinduced Vibrations Drive Ultrafast Structural Distortion in Lead Halide Perovskite}

\author{Hong-Guang Duan$^{1,2,3,*}$, Vandana Tiwari$^{1,4,*}$, Ajay Jha$^{1,*}$, Golibjon R. Berdiyorov$^{5}$, Alexey Akimov$^{6}$, Oriol Vendrell$^{7}$, Pabitra K. Nayak$^{8}$, Henry J. Snaith$^{8}$, Michael Thorwart$^{2,3}$, Zheng Li$^{1}$, Mohamed E. Madjet$^{5}$ \& R. J. Dwayne Miller$^{1,3,9}$} 
\makeatletter
\let\saved@includegraphics\includegraphics
\AtBeginDocument{\let\includegraphics\saved@includegraphics}
\renewenvironment*{figure}{\@float{figure}}{\end@float}
\makeatother

\begin{document}

\maketitle

\begin{affiliations}
\item Max Planck Institute for the Structure and Dynamics of Matter, Luruper
Chaussee 149, 22761, Hamburg, Germany
\item I.\ Institut f\"ur Theoretische Physik,  Universit\"at Hamburg,
Jungiusstra{\ss}e 9, 20355 Hamburg, Germany
\item The Hamburg Center for Ultrafast Imaging, Luruper Chaussee 149, 22761
Hamburg, Germany
\item Department of Chemistry, University of Hamburg, Martin-Luther-King Platz 6, 20146 Hamburg, Germany
\item Qatar Environment and Energy Research Institute, Hamad Bin Khalifa University, Qatar Foundation, P.O. Box 34110, Doha, Qatar
\item Department of Chemistry, State University of New York at Buffalo, Buffalo NY, 14260, USA
\item Physikalisch-Chemisches Institut, Universit\"at Heidelberg, Im Neuenheimer Feld 229, 69120 Heidelberg, Germany
\item Department of Physics, University of Oxford, Clarendon Laboratory, Parks Road, Oxford OX1 3PU, United Kingdom 
\item The Departments of Chemistry and Physics, University of Toronto, 80 St.
  George Street, Toronto Canada M5S 3H6
\\
$^*$These authors contributed equally to this work. \\ 
\centerline{\underline{\date{\bf \today}}}
\end{affiliations}

\begin{abstract} 

Organic-inorganic perovskites have shown great promise towards their application in opto-electronics. The success of this class of material is dictated by the complex interplay between various underlying microscopic phenomena. The structural dynamics of organic cations and the inorganic sublattice after photoexcitation is hypothesized to have a direct effect on the material properties, thereby affecting the overall device performance. Here, we use ultrafast heterodyne-detected two-dimensional (2D) electronic spectroscopy to reveal impulsively excited vibrational modes of methylammonium (MA) lead iodide perovskite, which drive the structural distortion after photoexcitation. The vibrational analysis of the measured data allows us to directly monitor the time evolution of the librational motion of the MA cation along with the vibrational coherences of inorganic sublattice. Wavelet analysis of the observed vibrational coherences uncovers the interplay between these two types of phonons. It reveals the coherent generation of the librational motion of the MA cation within $\sim$300 fs, which is complemented by the coherent evolution of the skeletal motion of the inorganic sublattice. To rationalize this observation, we have employed time-dependent density functional theory (TDDFT) to study the atomic motion of the MA cation and the inorganic sublattice during the process of photoexcitation. The TDDFT calculations support our experimental observations of the coherent generation of librational motions in the MA cation and highlight the importance of the anharmonic interaction between the MA cation and the inorganic sublattice. Capitalizing on this interaction, our advanced theoretical calculations predict the transfer of the photoinduced vibrational coherence from the MA cation to the inorganic sublattice, which drives the skeleton motion to form a polaronic state leading to long lifetimes of the charge carriers. Our study uncovers the interplay of the organic cation and inorganic sublattice during the formation of the polaron, which may lead to novel design principles for next generation perovskite solar cell materials.

\end{abstract} 

On the quest for a low cost and easily processable photoactive material, organic-inorganic
lead halide perovskites (LHPs) have emerged as a class of material with tremendous potential. Photovoltaics based on LHPs have displayed a remarkable increase in the power conversion efficiencies (PCE) as compared to other technologies in the past decade advancing to 24.2$\%$ based on a solid polycrystalline perovskite \cite{Nat_Rev_Mat_4_269_(2019), Prog_Phtovolt_Res_Appl_27_565_(2019), Science_348_1234_(2015), Nature_517_476_(2015), Energy_Environ_Sci_6_1739_(2013)}. Largly due to their easy solution processability, large carrier diffusion length and high photoluminescence quantum yield \cite{Nature_499_316_(2013), Science_342_341_(2013), Science_342_344_(2013)}, perovskites have also found their application in other areas such as photodetectors and lasing \cite{Science_348_1234_(2015), Nat_Mater_14_636_(2015), Science_338_643_(2012), Sci_Rep_2_591_(2012), JACS_134_17396_(2012), Nat_Photonics_9_687_(2015)}. Despite the success of the material to reach macroscopic performance scales such as the highly efficient PCE, a comprehensive understanding of the underlying microscopic phenomena of charge generation following photo-excitation is still far from being understood. 

To unravel the microscopic mechanisms of the success of perovskite in photoinduced charge generation, numerous experimental and theoretical works have been reported focusing on understanding the elementary photophysical processes in LHPs \cite{JPCL_9_6853_(2018), ACR_49_166_(2016), Sci_Adv_3_e1701469_2017, Adv_Mater_30_1707312_(2018), ACSPhoton_5_852_(2018), ACSPhoton_5_648_(2018), JACS_139_18262_(2017), JACS_140_9882_(2018)}. One of the important unresolved questions pertaining to perovskite photophysics is the origin of long charge carrier lifetimes using processing methods that generally lead to mid gap states.  Those act as efficient recombination centres, although the charge carrier mobilities in LHPs have been reported to be modest, i.e., only $\sim$50-100 cm$^{2}$V$^{-1}$s$^{-1}$ \cite{Science_342_341_(2013), JACS_136_13818_2014, EES_7_2269_2014}. This puzzling observation has attracted enormous attention by various research groups. To understand the underlying reasons for the long charge carrier lifetime in $\rm CH_3NH_3PbI_3$, Zhu and coworkers used a combination of the time-resolved optical Kerr effect and photoluminescence spectroscopy to compare the ultrafast hot-electron cooling dynamics in perovskites with different organic cations \cite{Science_353_1409_2016}. There, the reorientational motion of the MA cation has been identified to form the charge screening effect, which protects the energetic carriers via the formation of a large polaron on time scales competitive with that of ultrafast thermalization and recombination. In addition, based on the resonant and off-resonant excitations, the existence of a large polaron formation has been further supported in order to explain the modest charge mobilities of electrons in perovskite \cite{Sci_Adv_3_e1701469_2017, Sci_Adv_3_e1701217_2017}. The combination of the experiments and calculations has proposed that the deformation of the PbBr$^{-}_{3}$ sublattice is mainly responsible for the polaron formation. The intrinsic effects of organic cations on the band-edge charge carrier lifetime were studied by time-resolved photoluminescence of LHPs as a function of temperature \cite{PNAS_114_7519_2017}, which also supports the hypothesis of the formation of a large polaron. To capture the lattice displacement induced by photo-carrier generation in LHPs, recently LHPs have also been studied using impulsive vibrational spectroscopy \cite{Nat_Comm_9_1971_2018} and electron diffraction \cite{Sci_Adv_3_e1602388_2017}. Due to the proposed role of the organic cation in perovskite photo-physics, its vibrational dynamics has been extensively studied by various spectroscopic approaches \cite{JPCL_6_3663_2015, JACS_139_4068_2017, JPCL_9_895_2018, NANO_Lett_17_4151_2017}. Raman and photoluminescence spectroscopy have been used to investigate the structure-function relationships in MAPbI$_{3}$. The librational motion of the MA cation is found to be strongly coupled to the $\rm PbI_6$ perovskite octahedra by a hydrogen bond \cite{NANO_Lett_17_4151_2017}. Moreover, the vibrational dynamics of the inorganic sublattice have been studied during the process of the polaron formation \cite{Nat_Comm_9_2525_2018}. The observed low-frequency modes of the Pb-I bending and stretching vibrations result in the charge separation in the LHPs. Interestingly, resonant THz phonon excitation shows direct evidence of the mode-driving band gap in the LHPs \cite{Nat_Comm_8_687_2017}, which demonstrates the correlation of the band gap and the Pb-I-Pb angle bending vibrations. Advanced theoretical calculations have uncovered that the polaron formation is induced by the structural disorder resulting mainly from thermal distortions of the inorganic sublattice. They reduce the overlap between the electron and hole wave functions and the probability of bimolecular recombination is then lowered by two orders of magnitude \cite{Ener_Envi_Sci_11_101_2018}. In addition to the polaron formation, Rashba spin-orbit interaction has been considered to enhance the carrier lifetime in LHPs \cite{Nano_Lett_15_7794_2015}. On the basis of first-principle calculations and a Rashba spin-orbital model, Zheng and coworkers have suggested that the recombination rate is reduced due to spin-forbidden transitions. The subsequent transient absorption measurement is consistent with the existence of a giant Rashba splitting in 2D LHPs \cite{Sci_Adv_3_e1700704_2017}. This Rashba effect has also been reported in LHPs thin films \cite{Nat_Mat_16_115_2017}. In addition, the vibrational coherence of the librational motion of the MA cation has been calculated in a multi-unit-cell model of perovskite \cite{JPCL_6_693_2015}, showing that a ferroelectric domain wall is constructed to reduce the electron-hole recombination. Despite these enormous efforts, there has been no conclusive experimental report capturing the dynamics of the organic cation and its interaction with the inorganic sublattice during the formation of the large polaron. Such an analysis can reveal the degree to which the cation is involved and whether this potential microscopic mechanism for the long carrier lifetime is sufficiently strong.

To resolve the ultrafast structural dynamics of the polaron formation in ${\rm CH_3NH_3PbI_3}$, we study the coherent vibrational dynamics of the inorganic sublattice and its correlation with the librational motion of the organic cation using ultrafast two-dimensional electronic photo-echo spectroscopy (2DES). 2DES measures the third-order nonlinear response originating from the interaction of the sample with three resonant electric fields. The response is commonly plotted as a function of both the excitation and probe frequencies. In contrast to conventional pump-probe schemes, 2DES spectrally decomposes the optical signal into the  excitation and detection windows. By this, 2DES is able to disentangle the underlying transitions to provide the distinct selectivity of transient optical signatures and the transfer pathways. Due to these advantages, 2DES has recently been used to study the exciton dynamics and the carrier thermalization processes in LHPs \cite{ACSPhoton_5_852_(2018), Nat_Comm_8_376_2017, JPCL_8_3211_2017}. Here, we have judiciously tuned our laser spectrum to capture the motion of the nuclear wave packet in the electronically excited state by the spectrally disentangled excited-state absorption peak. 
The subsequent data analysis shows clear evidence of the coherent generation of the librational motion of the MA cation and the  skeletal motion of the inorganic sublattice. The subsequent wavelet analysis enables us to directly monitor the formation and the coherence of the  librational motion, resolved on an ultrafast timescale of 300 fs. In addition, it also reveals that the coherent generation of the MA cation is complemented by the motion of the inorganic sublattice in perovskite. Furthermore, we report advanced theoretical calculations to capture the coherent evolution of the librational modes of the MA cation. Our theory reveals a strong anharmonic interaction between the MA cation and the skeletal motion of the inorganic sublattice in pervoskite. This plays a significant role to drive the excited-state wave packet to form a polaron state. Based on this combination of experimental and theoretical efforts, we conclude that the interaction of the librational motion of the organic cation with the inorganic sublattice is responsible for  the formation of the polaronic state. 

\section*{Results} 
\begin{figure}
\begin{center}
\includegraphics[width=15cm]{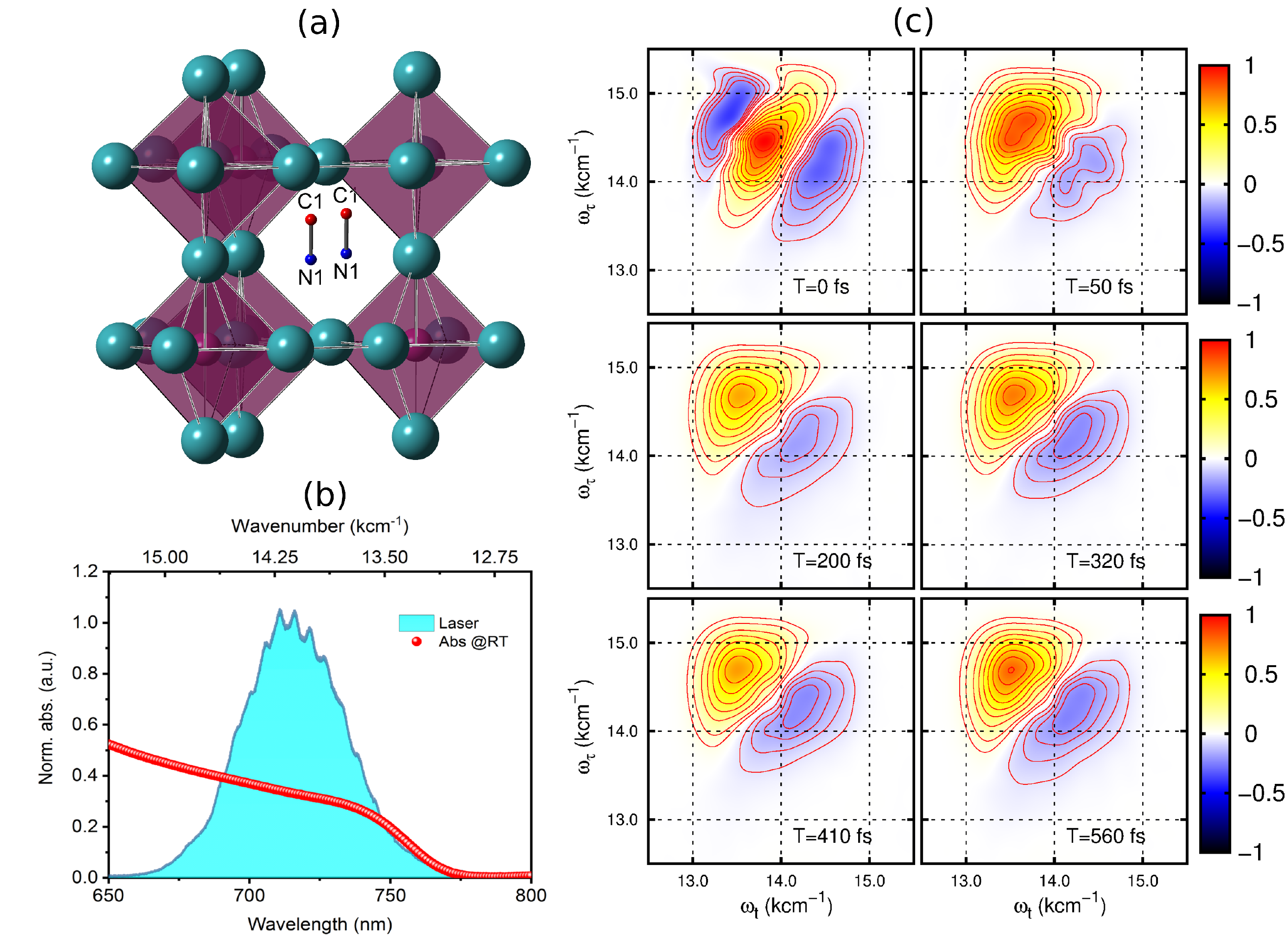}
\caption{\label{fig:Fig1} Ground state structure of tetragonal $\rm MAPbI_{3}$. (b) Steady-state absorption spectrum (red circle) of perovskite at room temperature and laser spectrum (light blue area). (c) Time evolution of 2D electronic spectra (real part) at the selected waiting times. The magnitude of the spectra decays with growing waiting time. Interestingly, in 2D spectra, the oscillation of the  amplitude from 200 fs to 560 fs can be clearly observed. Positive and negative amplitudes indicate the ground state bleach and excited state absorption, respectively. }
\end{center}
\end{figure}

The solution-processed $\rm CH_3NH_3PbI_3$ is prepared (details are described in the section of Materials and Methods) and grown on a quartz substrate with 1 mm thickness. For the optical measurement, the sample film is mounted in the cryostat (Oxford Instrument) with vacuum condition to avoid the degradation by moisture. Fig.\ \ref{fig:Fig1}(a) shows the molecular structure of tetragonal $\rm CH_3NH_3 PbI_3$ with the MA cation. In addition, the steady-state absorption spectrum of $\rm CH_3NH_3PbI_3$ is shown in Fig.\ \ref{fig:Fig1}(b). The laser spectrum used in the present measurements is marked by the light-blue shaded area. 

\subsection{Two-dimensional electronic photon echo spectroscopy}
To study the exciton and free carrier dynamics after photoexcitation, we measure the 2D electronic spectra of $\rm CH_3NH_3PbI_3$ at room temperature (296 K). The details of our home-built 2D spectrometer and the acquisition procedures are described in the Materials and Methods section. The real part of the 2D spectra at selected
waiting times are shown in Fig.\ \ref{fig:Fig1}(c). At T = 0 fs, the peak in the 2D spectrum shows  clear
evidence of elongation along its diagonal, which indicates the presence of strong inhomogeneous broadening. Moreover, the strong absorption features at zero waiting time (T = 0 fs) overlap with the central (positive) peak, which induces a dramatic shift of the central peak to the upper left. At T = 50 fs, the magnitude of the 2D spectrum is strongly reduced and the peaks show less elongation along the diagonal, which indicates a rapid reduction of the strong inhomogeneous broadening within this initial time range. The inhomogeneity of the peaks is hard to be observed at T = 200 fs and the amplitude of the peaks is further reduced compared to the one at T = 50 fs. The initial dynamics of the photoexcited perovskite within 200 - 300 fs have been attributed to lattice reorganization and carrier thermalization processes \cite{Nat_Comm_8_376_2017, JPCL_8_3211_2017}. The shape of the peaks does not significantly change in the 2D spectra with increasing waiting time after 200 fs. Interestingly, the magnitude of the peaks in the 2D spectra at 320 fs is larger than the one at 200 fs and the same behaviour is observed from 410 fs to 560 fs, which clearly illustrates the presence of oscillations with large amplitude in the 2D electronic spectra. 

\subsection{Coherent Vibrational Dynamics} 

\begin{figure}
\begin{center}
\includegraphics[width=12cm]{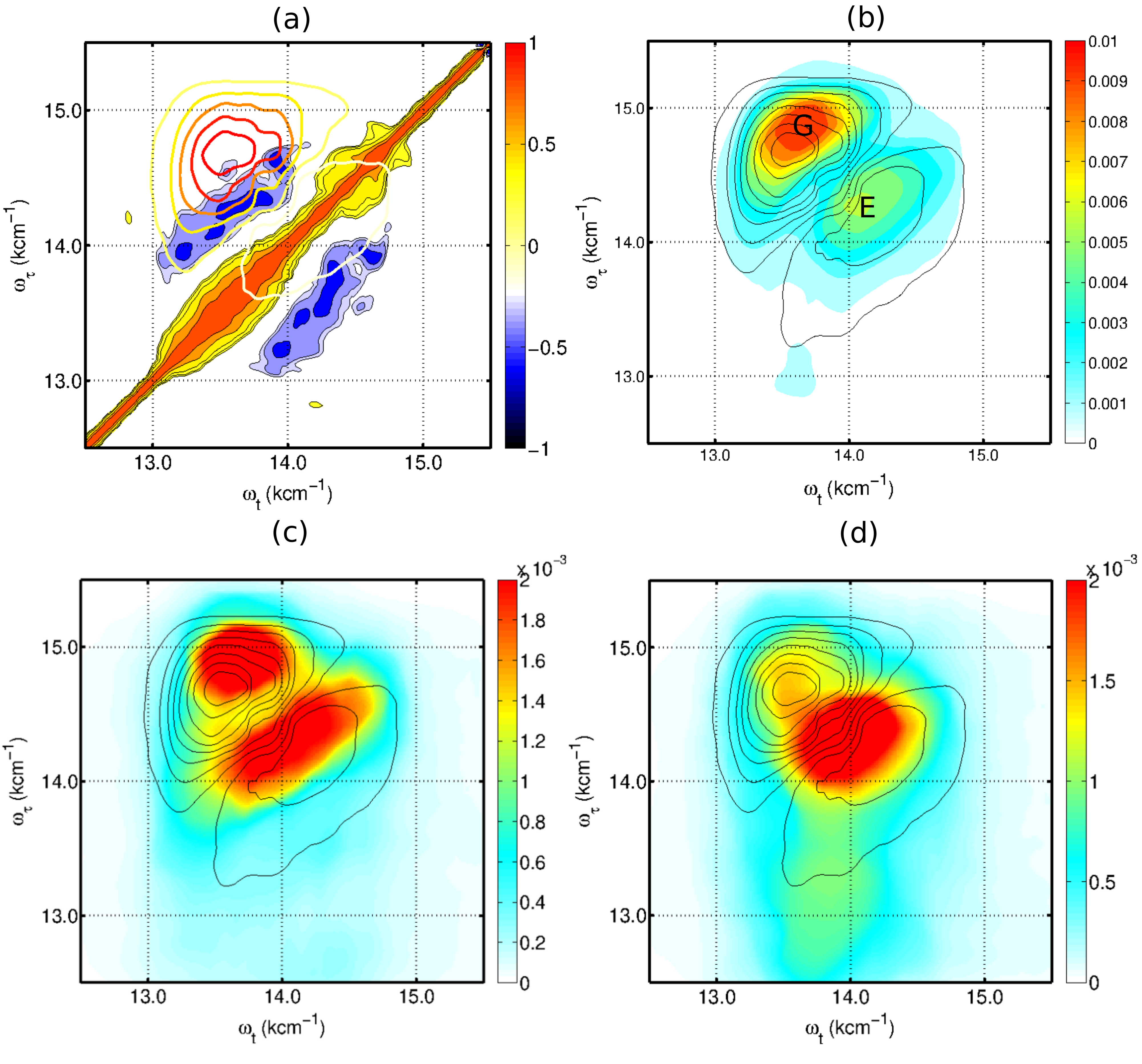}
\caption{\label{fig:Fig2} (a) 2D correlation spectrum after analyzing correlated cross peaks along diagonal. 2D electronic spectrum at T=200 fs is shown as contours for comparison. (b)-(d) 2D power spectra at frequencies of 157, 65 and 81 cm$^{-1}$, respectively. For comparison, 2D spectrum at T=200 fs is plotted as black contours. For data analysis, we extract the kinetics of selected peak at G and E in (b) to examine their coherent dynamics in GSB and ESA, respectively. } 
\end{center}
\end{figure}

To examine the oscillations observed in 2DES, we first collect a three-dimensional (3D) set of data consisting of 2D spectra with varying waiting times T. To retrieve the oscillatory signal of the 2D spectra, we perform a global fitting by exponential functions on the 3D data and analyze the oscillations along the excitation and detection frequencies of the residuals obtained after removing the global kinetics. To assign the origin of the observed oscillations, we perform a 2D correlation analysis of the 2D residual maps, in which the oscillations of the correlated two peaks are analyzed along the diagonal direction in the 2DES. The resulting 2D correlation map is shown in Fig.\ \ref{fig:Fig2}(a) and the details of this analysis are given in the SI. We observe negative peaks with a strong amplitude which are present along the diagonal in the 2D correlation spectrum. These negative peaks provide strong evidence of anti-correlated oscillations at the presented coordinates in the 2DES. It has been demonstrated that anti-correlated oscillations in 2D spectra originate from underlying vibrational coherences \cite{CPL_545_40_2012, JPCA_118_10259_2014}. Thus, this correlation analysis uncovers the vibrational origin of the coherence in the observed oscillations in the 2DES data. 

\begin{figure}
\begin{center}
\includegraphics[width=16cm]{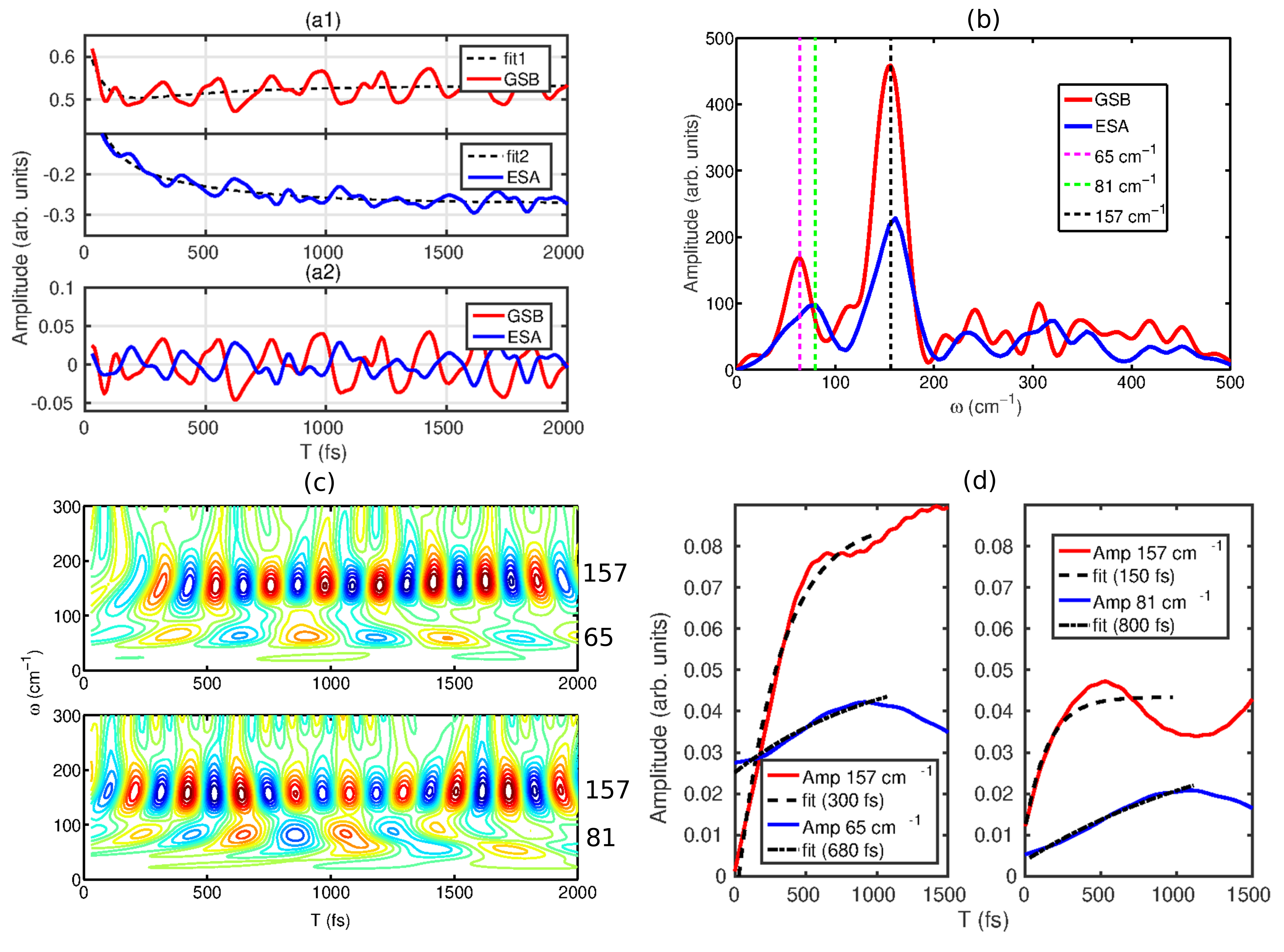}
\caption{\label{fig:Fig3} (a1) traces at GSB (red solid line) and ESA (blue solid line). Their coordinates are marked as G and E separately in Fig.\ \ref{fig:Fig2}(b). These traces are fitted by global fitting approach in (a1) and the residuals are obtained after removing kinetics. In (a2), the polished residuals are shown as red (G) and blue (E) solid lines after removing the high-frequency noise. (b) The identified vibrational frequencies in power spectrum. The red (G) and blue (E) solid lines show the vibrational modes at 65, 81 and 157 cm$^{-1}$. (c) Wavelet analysis of residuals on GSB and ESA. The coherent dynamics of 65, 81 and 157 cm$^{-1}$ modes from  GSB (upper panel) and ESA (lower panel) are presented at the left and right part in (d). In GSB (left panel), coherent generation of 65 and 157 cm$^{-1}$ shows an increase in amplitude on the timescale of 680 and 300 fs, respectively. The coherent dynamics of modes at 81 and 157 cm$^{-1}$ show the amplitude increase on timescales of 800 and 150 fs in the ESA contribution (right panel), respectively.} 
\end{center}
\end{figure}

Next, we analyze the frequencies of these vibrational oscillations by employing a Fourier transform of the 2D residuals obtained after removing the kinetics by the  global fitting approach. We identify a few key modes with the largest amplitudes and plot the absolute values of the observed distinct frequencies along $\omega_{\tau}$ and $\omega_{t}$ in 2D power spectra. We show  selected 2D power spectra for the frequencies of 157, 65, 81 cm$^{-1}$ in  the panels (b) to (d) of Fig.\ \ref{fig:Fig2}. In addition, two retrieved 2D power spectra of the modes at 33 and 48 cm$^{-1}$ are shown in the SI. The black contour of the 2D spectrum at 200 fs is overlayed with the 2D power maps in order to identify the location of the oscillations. This comparison reveals that the amplitude of the 2D power map at 157 cm$^{-1}$ perfectly overlays with the ground-state bleach (GSB) and the excited-state absorption (ESA) features in the 2D electronic spectrum. The vibrational mode of 157 cm$^{-1}$ shows the strongest magnitude in the 2D power spectra with the time period of 208 fs, which perfectly agrees with the amplitude fluctuation observed in Fig.\ \ref{fig:Fig1}(c). The strong amplitude in Fig.\ \ref{fig:Fig2}(b) shows  clear evidence of strong vibronic coupling of this key mode to the electronic transitions in our 2DES. Moreover, in this 2D power spectrum, we observe two separated peaks, marked by G and E, which are connected by a node. It implies that the dynamical oscillations at these two peaks have anti-correlated phases, which agrees with the observations of the 2D correlation analysis in Fig.\ \ref{fig:Fig2}(a). In Fig.\ \ref{fig:Fig2}(c)-(d), we clearly observe  that the magnitudes of vibrations are mainly distributed in the GSB and ESA regions with the separation of a node. To further explore the origin of this coherence, we extract the kinetical traces at the maximum amplitude of the G and E peaks in Fig.\ \ref{fig:Fig2}(b) and plot them as red (G) and blue (E) solid lines in Fig.\ \ref{fig:Fig3}(a1). To retrieve residuals, we use the global fitting approach to fit their kinetics and show the result of the fit as black dashed lines. The residuals of the raw data are shown as red and blue dashed lines in the lower part of Fig.\ S6 in the SI. Moreover, the high-frequency noise is removed by a tukey-window Fourier transform and the polished residuals are presented as red and blue solid lines for the GSB and ESA in Fig.\ \ref{fig:Fig3}(a2), respectively. We observe that the phases of the oscillations shown by the red and blue lines are anti-correlated which demonstrates the validity of our analysis reported in the 2D correlation spectrum in Fig.\ \ref{fig:Fig2}(a). The details of the tukey-window Fourier transform are described in the SI. 

To distinguish the vibrations of the electronic ground or excited states, we construct a theoretical model with three electronic states and identify the optical signals of the GSB and ESA by assigning the optical transitions from the ground to the first excited state and the transition from the first to the second excited state, respectively. The 2D electronic spectra are calculated for different waiting times and the retrieved vibrational coherences from the GSB and ESA peaks are further analyzed to extract the phase relation. More details on the calculations of the 2D spectra are described in the SI. Based on our theoretical calculations, the anti-correlated oscillations from the GSB and ESA demonstrate the validity of our model and shows the separation of vibrations from the electronic ground and excited states (see Figure S3 in the SI). By this, we are able to monitor the electronic and vibrational dynamics from the ground and excited states by tracking the kinetics of the GSB and ESA peaks. To retrieve their oscillation frequencies, we perform the Fourier transform of the two residuals and plot the power spectra in Fig.\ \ref{fig:Fig3}(b). We clearly identify three key modes at the frequencies of 65, 81 and 157 cm$^{-1}$, which are marked by the magenta, green and black dashed lines, respectively. Due to the lifetime broadening, the low-frequency modes at 33 and 48 cm$^{-1}$ are not well resolved in Fig.\ \ref{fig:Fig3}(b). It shows a strong mode at 157 cm$^{-1}$ both in the GSB and the ESA. However, in the region of  low frequency ($<$100 cm$^{-1}$),   slightly different frequencies in the GSB and ESA are found. To obtain the lifetime of this coherent generation, we analyze the data from the GSB and ESA region using the wavelet analysis. The spectra obtained are shown in Fig.\ \ref{fig:Fig3}(c) (details of the analysis are described in the SI). We extract the kinetics of the oscillations at 65 and 157 cm$^{-1}$ and obtain the time evolution of the amplitude by fitting of sine function with exponential decay. The time-evolved amplitudes are shown in the left part of Fig.\ \ref{fig:Fig3}(d). The amplitudes for the cases 65 and 157 cm$^{-1}$ are plotted as blue and red solid lines, respectively. The exponential fit of the data reveals a coherent generation of vibrations at 157 cm$^{-1}$ within 300 fs. Using a similar analysis, we observe the enhancement of the vibrational coherence of 65 cm$^{-1}$ with a timescale of 680 fs. In addition, we perform the same data analysis for the traces of 81 and 157 cm$^{-1}$ in the ESA part and plot the results in the right panel of Fig.\ \ref{fig:Fig3}(d). We find a coherent generation of vibrations at 81 and 157 cm$^{-1}$ within 800 and 150 fs, respectively. 

\subsection{Computational Analysis} 

To identify the origin of the experimentally observed vibrational coherences and to simulate their dynamics, we have performed theoretical calculations of the photoexcitation process in $\rm CH_3NH_3PbI_3$ (details of the computations are provided in the Methods section). First, we perform DFT calculations in the tetragonal phase of perovskite. Fig.\ \ref{fig:Fig4}(a) represents the difference of electron localization functions for the ground and the excited states (the excited state has been simulated by adding 10$^{19}$ cm$^{-3}$ excess electrons). As we show in Fig.\ \ref{fig:Fig4}(a), the electrons are mainly localized on the Pb atom and this excess charge strongly interacts with the positively charged NH$^{+}_{3}$ group. It forces the dipole of the MA cation to rotate towards the Pb atom (this is supported by the argument elaborated in the Discussion Section). To capture the subsequent structural dynamics, we carry out the excited-state molecular-dynamics calculation for a model system with ${2}\times{2}\times{2}$ supercells for the time evolution of 2 ps. We show the dominant principal axis as the representative motion of the model system in Fig.\ \ref{fig:Fig4}(b) to (e), in which librational motions of the MA cation at 158 cm$^{-1}$ and the Pb-I lattice distortion appear as the major components of the dynamical processes. Specified configurations of the skeletal vibrations are shown in Fig.\ \ref{fig:Fig4}(c) to (d) with the frequencies of 61 and 87 cm$^{-1}$, respectively. In addition, our simulations show two low-frequency modes of 37 and 47 cm$^{-1}$, they are described in the SI. Moreover, our dynamical calculations reveal the coherent generation of the librational motion of the MA cation at 158 cm$^{-1}$. In Fig.\ \ref{fig:Fig4}(e), we observe that the magnitude of the vibration at 158 cm$^{-1}$ (marked by black dashed line) gradually increases with evolving time and reaches its maximum at 300 fs. This timescale of the coherent generation perfectly agrees with the timescale revealed by the experimental observations reported above. In addition, our theoretical calculations uncover an anharmonic interaction between the strong coherent oscillations of the MA cation and the skeletal motion of the inorganic sublattice in perovskite. Induced by this interaction, the vibrational coherences of the skeletal motion is strongly modulated during the coherent generation of the librational motion of the MA cation, which is clearly shown in Fig.\ \ref{fig:Fig4}(e). These structure-based calculations allow us to identify the atomic motions of the MA cation, which strongly interact with the adjacent skeletal (I-Pb-I) motions after being subject to the optical transition from the valance to the conduction band. Thus, along with the librational motion of the MA cation, the calculated vibrational motions corresponding to the  inorganic sublattice observed at 37, 47, 61 and 87 cm$^{-1}$ actively participate in the process of the polaron formation. 

\subsection{Discussion} 

Based on the experimental and computational approaches described above, we conclude that the electron-sublattice interaction in the photoexcited tetragonal $\rm CH_3NH_3PbI_3$ leads to the formation of a large polaron and involves the interplay of different vibrational frequencies representing organic and inorganic sublattices’ motion. Specifically, the intense vibrational coherences of the methyl librational mode at 158 cm$^{-1}$ dominate the initial dynamics at early times of 200-300 fs. In order to clarify the underlying mechanism for the build-up of the librational amplitude of the MA cation, we have analyzed the temporal evolution of the difference of the electronic density between the excited and the ground states, $\rm \Delta \rho(t) = \rho_{ES}(t)- \rho_{GS}(t)$, for the $\rm CH_3NH_3PbI_3$ model system subject to the same excited state structures in the MD trajectory. As shown in Fig.\ \ref{fig:Fig4}(f), both the initially created electron and hole are delocalized subject to the perovskite structure close to the tetragonal symmetric equilibrium ground state. The localization of the excess electron on the Pb atom aligns the MA cation by rotating its molecular axis. This generates the strong signal of the vibrational coherence of the MA cation on an ultrafast timescale of 300 fs. The strong interaction transfers the coherence from the librational motion of the organic cation to the motion of the sublattice of the I-Pb-I skeleton. Based on our simulations, we find that this transfer of coherence is finished within a time span of 1 ps, which can  be observed  in Fig.\ \ref{fig:Fig4}(f) at 700 fs. We show the detailed time evolution of electronic charge on skeleton of perovskite in the SI (Fig.\ S7). The localization of the charge density, in turn, enhances the MA librational amplitude. With this redistributed charge, the skeletal motion of $\rm CH_3NH_3PbI_3$ slightly changes its frequencies to reach the (metastable) local potential minimum of the stabilized polaron.  

It is important to mention that recent studies on the excited state of LHP focus only on the skeletal modes that are comprised of the motion of the inorganic sublattice \cite{Nat_Comm_9_1971_2018, Nat_Comm_9_2525_2018}, but the information on the dynamics of the organic cation has not been given its due importance. Nevertheless, it could be that, in the ground state, the organic cation is electronically decoupled from the inorganic sublattice, as is shown by a study using an infrared-pump electronic-probe measurement \cite{Nat_Comm_10_482_2019}. However, after photoexcitation, the reorganization of the lattice due to the polaron formation leads to a significant interaction between the organic cation and the PbI$^{-}_{3}$-sublattice. In fact, this interaction provides the necessary stabilization for the polaronic state that effectively localizes the carrier by reducing the electron-hole carrier overlap which blocks carrier recombination. The resulting electrostatic screening, unique to the strong ionic character of LHPs, greatly distinguishes this class of materials from other semiconductors. It is this ultrafast intense polaron formation that is responsible for the observed long carrier lifetime in these materials. This holds even for simple methods of the sample preparation that are normally associated with high defect densities and rapid recombination losses. This class of materials is distinct from other polar and ionic semiconductors in that the structure comprised of an inorganic and a labile organic ionic sublattice leads to a localized charge density and an exceptionally strong coupling to the lattice and the noted ultrafast large polaron formation. 

\section*{Conclusions} 

We have studied the coherent dynamics of photoexcitations in tetragonal $\rm CH_3NH_3PbI_3$ at room temperature. Based on the high sensitivity of heterodyne-detected 2D electronic spectroscopy, we provide evidence of quantum coherences, which emerges after  photo-excitation of an electron from the valence band to the hot free-carrier band.  A 2D correlation analysis reveals the origin of the vibrational coherences in 2DES. Based on the phase analysis on the GSB and ESA peaks, the dominating modes with frequencies of 65, 81 and 157 cm$^{-1}$, which are coupled to the electronic states, are identified. Moreover,  the strong magnitude of the  oscillations manifests a strong vibronic coupling of these key modes to the electronic transitions during photoexcitation. Based on a wavelet analysis, we further reveal that the key mode of 157 cm$^{-1}$ is gradually generated on an ultrafast timescale of 300 fs. The amplitude of the low-frequency modes at 65 and 81 cm$^{-1}$ are enhanced during the process of the coherent generation of the 157 cm$^{-1}$ mode. Supported by our advanced theoretical calculations, we are able to assign the key mode of 157 cm$^{-1}$ to the librational motion of the MA cation and our theoretical prediction of the coherent generation and increasing amplitude the cation  motion occuring within the timescale of 300 fs. This perfectly agrees with our experimental observation. Moreover, we have identified the modes of 33, 48, 65 and 81 cm$^{-1}$ as the key modes related to vibrational coherence of the skeletal motion of PbI$^{-}_{3}$. A structure-based approach allows us to capture evidence of a strong interaction between the  MA cation and the inorganic sublattice. This interaction induces a coherent transfer of excitation from the librational motion of the MA cation to the skeletal motion of PbI$^{-}_{3}$, in which the sublattice motion of PbI$^{-}_{3}$ strongly modulates the photogenerated wave packet dynamics on the surface of the  electronic excited states. Moreover, we have theoretically captured the wave-packet dynamics which assists the polaron  formation by localizing the electron density on the skeletal structure of the Pb atom. Thus, our study provides a new comprehensive  understanding of the wave packet motion, which induces the vibrationally coherent polaron formation process, and hence unravels the underlying mechanism of the charge generation and protection in $\rm CH_3 NH_3PbI_3$. Based on this finding, one may anticipate rational design principles for the future improvement of perovskite materials. 

\begin{figure}
\begin{center}
\includegraphics[width=10cm]{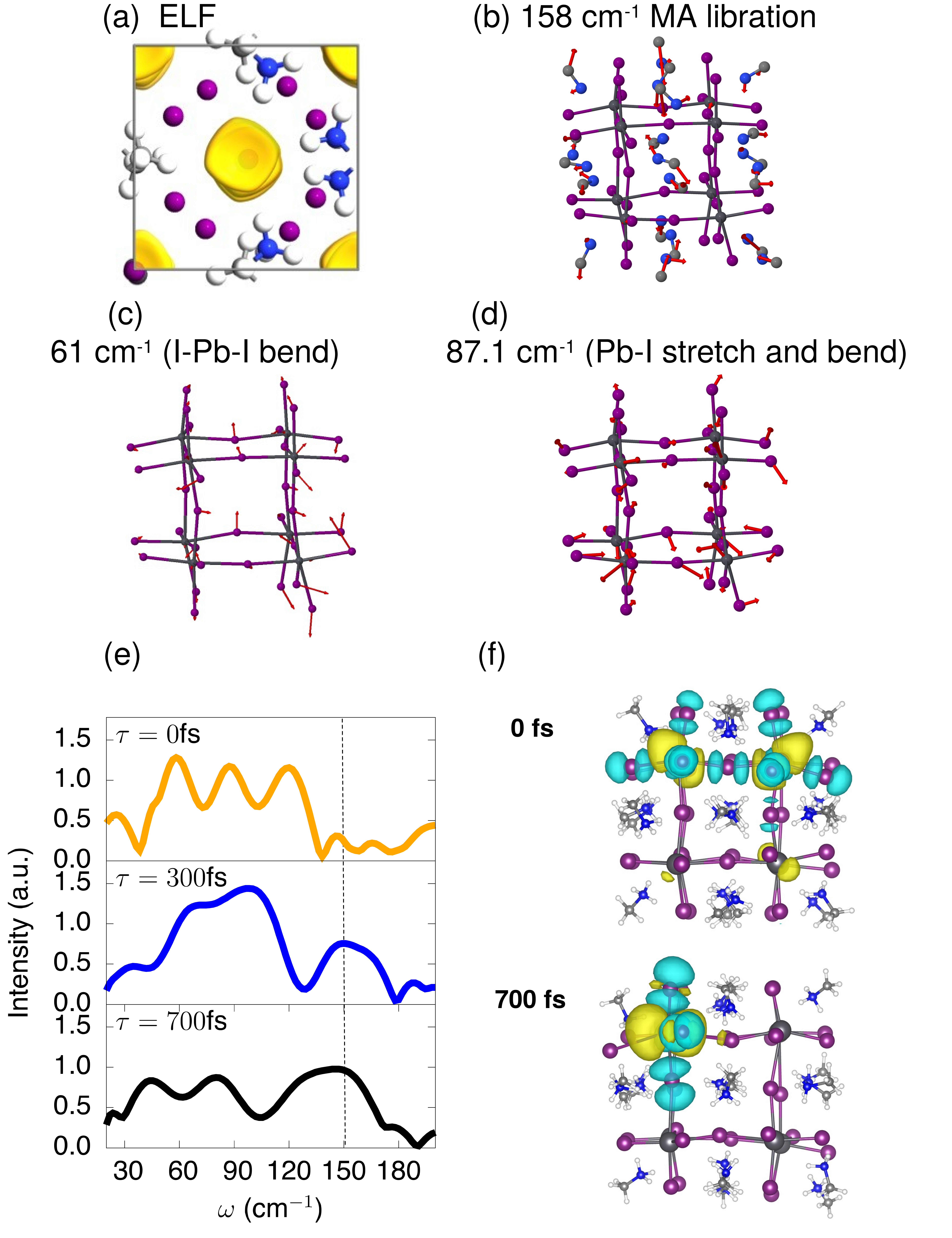}
\caption{\label{fig:Fig4} (a) Difference of electron localization functions with 10$^{19}$ cm$^{-3}$ excess electrons compared to its ground state. The magnitude of its isosurface is 0.03. (b) Molecular configuration of perovskite with the librational motions of the MA cation at 158 cm$^{-1}$. The vibrational coherence of the inorganic sublattice is present in the low-frequency region, 61 and 87 cm$^{-1}$, which are shown in (c) to (d), respectively. (e) Time evolution of the vibrational coherence after photoexcitation. The librational motion of the MA cation (158 cm$^{-1}$) is generated on a  timescale of 300 fs. Moreover, the low-frequency ($<$100 cm$^{-1}$) shift is presented within $\sim$700 fs after photoexcitation. (f) Based on the ground-state equilibrium structure, the initial electron density is strongly delocalized after photoexcitation. The electron density is localized after reaching equilibrium at 700 fs. The MA cation rotates its molecular axis to align to the localized charge on the Pb atom. In addition, the sublattice is slightly distorted compared to the equilibrium configuration of ground state. } 
\end{center}
\end{figure}

\begin{methods}
\subsection{Experimental setup} 

Ultrashort coherent pulses were generated by a home-built nonlinear optical parametric amplifier pumped by a commercial femtosecond laser Pharos (Light Conversion). A broadband spectrum with a linewidth of 100 nm (FWHM) was centred at 13800 cm$^{-1}$ such that an overlap with the near infrared region of the absorption spectrum of Perovskites is achieved. The excitation pulse was further compressed to the Fourier transform-limit with duration of 16 fs by the combination of a prism pair (F2) and a deformable mirror (OKO Technologies). Their temporal profiles were characterized by means of frequency-resolved optical grating (FROG). The measured FROG traces were analyzed using a commercial program FROG3 (Femtosecond Technologies). The 2D spectra were collected in an all-reflective 2D spectrometer based on a diffractive optic (Holoeye) with a phase stability of $\lambda/160$ whose configuration is described elsewhere \cite{SPP_162_432_(2015)}. Further components were the Sciencetech spectrometer model 9055 and a high-sensitive CCD linear array camera (Entwicklungsb{\"u}ro Stresing). The 2D spectra were collected at each fixed waiting time T by scanning the delay time $\tau = t_{1}-t_{2}$ in the range of [-128fs, 128fs] with a delay step of 1 fs. At each delay point, 200 spectra were averaged to reduce the noise ratio. The waiting time $T = t_3-t_2$ was linearly scanned in the range of 0 - 2 ps in steps of 10 fs. The energy of the excitation pulse is limited to 5 nJ with 1 KHz repetition rates for room temperature measurements. To check the pump dependence, we have also carried out measurements with different pump energies. Three pulses are focused on the sample with the spot size 100 $\mu$m and the photon echo signal generated at the phase-matching direction. To avoid oxidation, the sample was kept in the cryostat under vacuum condition ($\rm 1.7\times10^{-7}$ mPa) at room temperature. To verify the reproducibility of the results, measurements are performed on different spots on the perovskite films as well as with films prepared from three different batches. All showed similar spectral features. 

\subsection{Sample preparation and measuring condition}

Acetonitrile (ACN), gamma butyrolactone (GBL) Methyl ammonium solution, Chloroform and PbI2 were procured from Sigma Aldrich and used as received. MAI was procured from Dyesol.  Tetragonal single crystals of MAPbI3 were prepared by a  method described in Ref.\ \cite{Chem_Soc_Jap_45_1030_2016} and were used as a precursor material. The precursor solution was prepared by dissolving single crystals of MAPbI3 in an acetonitrile and methyl amine solution \cite{Ener_Envi_Sci_10_145_2017}. The final concentration of the solution is 0.5 M. The perovskite films were prepared by spin-coating the precursor solution at 2000 r.p.m on quartz substrates for 45 second in a $\rm N_2$ purged dry box. The UV-Vis spectrum of the thin film was taken on a Carry 300 UV-Vis spectrometer.

\subsection{Theoretical model} 

The density functional theory calculations with periodic boundary conditions are performed using \texttt{QuantumATK}. To mimic the charge localization on the Pb atom, electrons with a density of $10^{19}~\text{cm}^{-3}$ have been injected into the system. By this, we can study the motion of electrons in the electronically excited state of perovskite \cite{Nano_Lett_16_3809_2016}. The characterization of the localized charge in Fig.\ \ref{fig:Fig4}(a) is done by the electron localization function \cite{JCP_92_5397_1990}, which quantifies the probability of finding a second like-spin electron in the vicinity of an electron at a given space point. 

The Born-Oppenheimer excited state molecular dynamics has been computed for a model system with ${2}\times{2}\times{2}$ supercells for 2 ps, on the configuration interaction singles (CIS) level with the SBKJC effective core potential \cite{JCP_81_6026_1984}, using Chemical Dynamics Toolkit (CDTK) \cite{JCP_138_094311_2013} and GAMESS \cite{JCC_14_1347_1993}. The MD trajectories are initiated from the structures in the ground state tetragonal (I4/mcm) geometry after equilibration at room temperature. Its lowest unoccupied orbital is highly localized on a specific Pb atom, which agrees with the results from PBC-DFT. We have analyzed the excited state MD trajectories by their principal component \cite{Science_350_1501_2015}, which has been obtained by diagonalizing the covariance matrix $C=\langle{\bf x}{\bf x}^{T}\rangle$. Here, ${\bf x}={\bf r}-\langle\bf {r}\rangle$ are the atomic displacement vectors in the 3N dimensional configuration space for structures in the MD trajectories. To obtain the dynamics of the dominant principal motions, we project the vibrations onto the principal axes to show the substantial structural transition from the initial geometry to the excited state polaronic geometry within a time of 1 ps. Because the skeletal modes of the Pb-I lattice ($<$100 cm$^{-1}$) are similar in the ground and excited state \cite{Nat_Comm_9_2525_2018}, we decompose this principal axis vector of the atomic displacements with vibrational eigenvectors of the ground state perovskite by taking their inner products. We obtain the vibrational spectrum from the time evolution of the atomic velocities $\vec{v}_k(t)$ in the excited state, with consideration of the large amplitude motion \cite{JPCA_108_11056_2004,PCCP_15_6608_2013,JPCB_119_8080_2015} 
\begin{eqnarray} 
 \label{eq:spectrum} 
 &&I(\tau,\omega)= \mathcal{F}_t[I(\tau,t)](\omega)\nonumber \\
 &=&\mathcal{F}_t\left\{ \frac{1}{N_{at}}\sum_{k}^{N_{at}}\left\langle
      \textbf{v}_k(\tau)\cdot\textbf{v}_k(\tau +t)\exp(-\alpha t^2)
    \right\rangle \right\}(\omega)  
\,, 
\end{eqnarray} 
where $\mathcal{F}_t$ denotes the Fourier transform, $N_{at}$ is the number of atoms and the parameter $\alpha$ is chosen to suppress the autocorrelation function to effectively select the vibrational modes that are present around $\tau$. In this way, the temporal evolution of vibrational modes can be revealed by both their frequencies and amplitudes \cite{PCCP_15_6608_2013,JPCB_119_8080_2015}.    

\end{methods}

\bibliographystyle{naturemag} 

\begin{thebibliography}{11}
\bibitem {Nat_Rev_Mat_4_269_(2019)} Nayak, P. K., Mahesh, S., Snaith, H. J., \& Cahen, D. Photovoltaic solar cell technologies: analysing the state of the art. {\em Nat. Rev. Mat.} \textbf{4}, 269-285 (2019).
\bibitem {Prog_Phtovolt_Res_Appl_27_565_(2019)} Green, M. A., Dunlop, E. D., Levi, D. H., Hohl-Ebinger, J., Yoshita, M. \& Ho-Baillie, A. W. Y. Solar cell efficiency tables (version 54). {\em Prog. Photovolt. Res. Appl.} \textbf{27}, 565-575 (2019).
\bibitem {Science_348_1234_(2015)} Yang, W. S. {\em et al.} High-performance photovoltaic perovskite layers fabricated through intramolecular exchange. {\em Science} \textbf{348}, 1234-1237 (2015). 
\bibitem {Nature_517_476_(2015)} Jeon, N. J. {\em et al.} Compositional engineering of perovskite materials for high-performance solar cells. {\em Nature} \textbf{517}, 476-480 (2015).
\bibitem {efficiency} For record cell efficiencies, see: Best Research-Cell Efficiencies. http://www.nrel.gov/ncpv/images/efficiency-chart.jpg (2016).
\bibitem {Energy_Environ_Sci_6_1739_(2013)} Ball, J. M., Lee, M. M., Heya, A. \& Snaith, H. J. Low-temperature processed meso-superstructured to thin-film perovskite solar cells. {\em Ener. Envir. Sci.} \textbf{6}, 1739-1743 (2013).
\bibitem {Nature_499_316_(2013)} Burschka, J. {\em et al.} Sequential deposition as a route to high-performance perovskite-sensitized solar cells. {\em Nature} \textbf{499}, 316-319 (2013).
\bibitem {Science_342_341_(2013)} Stranks, S. D. {\em et al.} Electron-hole diffusion lengths exceeding 1 micrometer in an organometal trihalide perovskite absorber. {\em Science} \textbf{342}, 341-344 (2013).
\bibitem {Science_342_344_(2013)} Xing, G. {\em et al.} Long-range balanced electron- and hole-transport lengths in organic-inorganic ${\rm CH_{3}NH_{3}PbI_{3}}$. {\em Science} \textbf{342}, 344-347 (2013).
\bibitem {Nat_Mater_14_636_(2015)} Zhu, H. {\em et al.} Lead halide perovskite nanowire lasers with low lasing thresholds and high quality factors. {\em Nat. Mater.} \textbf{14}, 636-642 (2015).
\bibitem {Science_338_643_(2012)} Lee, M. M., Teuscher, J., Miyasaka, T., Murakami, T. N. \& Snaith, H. J. Efficient hybrid solar cells based on meso-superstructured organometal halide perovskites. {\em Science} \textbf{338}, 643-647 (2012).
\bibitem {Sci_Rep_2_591_(2012)} Kim, H. -S. {\em et al.} Lead iodide perovskite sensitized all-solid-state submicron thin film mesoscopic solar cell with efficiency exceeding 9\%. {\em Sci. Rep.} \textbf{2}, 591 (2012).
\bibitem {JACS_134_17396_(2012)} Etgar, L. {\em et al.} Mesoscopic ${\rm CH_{3}NH_3PbI_3/TiO_2}$ heterojunction solar cells {\em J. Am. Chem. Soc.} \textbf{134}, 17396-17399 (2012).
\bibitem {Nat_Photonics_9_687_(2015)} Lin, Q., Armin, A., Burn, P. L. \& Meredith, P. Filterless narrowband visible photodetectors. {\em Nat. Photonics} \textbf{9}, 687-694 (2015).

\bibitem {JPCL_9_6853_(2018)} Herz, L. M. How lattice dynamics moderate the electronic properties of metal-halide perovskites. {\em J. Phys. Chem. Lett.} \textbf{9}, 6853-6863 (2018).
\bibitem {ACR_49_166_(2016)} Saba, M., Quochi, F., Mura, A. \& Bongiovanni, G. Excited state properties of hybrid perovskites. {\em Acc. Chem. Res.} \textbf{49}, 166-173 (2016).
\bibitem {Sci_Adv_3_e1701469_2017} Miyata, K. {\em et al.} Lead halide perovskites: crystal-liquid duality, phonon glass electron crystals, and large polaron formation. {\em Sci. Adv.} \textbf{3}, e1701469 (2017). 
\bibitem {Adv_Mater_30_1707312_(2018)} Bretschneider, S. A., Ivanov, I., Wang, H. I., Miyata, K., Zhu, X. \& Bonn M. Quantifying polaron formation and charge carrier cooling in lead-iodide perovskites. {\em Adv. Mater.} \textbf{30}, 1707312 (2018).
\bibitem {ACSPhoton_5_852_(2018)} Jha, A., Duan, H.-G., Tiwari, V., Nayak, P. K., Snaith, H. J., Thorwart, M. \& Miller, R. J. D. Direct observation of ultrafast exciton dissociation in lead iodide perovskite by 2D electronic spectroscopy. {\em ACS Photonics} \textbf{5}, 852-860 (2018).
\bibitem {ACSPhoton_5_648_(2018)} Bohn, B. J., Simon, T., Gramlich, M., Richter, A. F., Polavarapu, L., Urban, A. S. \& Feldmann J. Dephasing and quantum beating of excitons in methylammonium lead iodide perovskite nanoplatelets. {\em ACS Photonics} \textbf{5}, 648-654 (2018).
\bibitem {JACS_139_18262_(2017)} Ghosh, T., Aharon, S., Etgar, L., \& Ruhman, S. Free carrier emergence and onset of electron-phonon coupling in methylammonium lead halide perovskite films. {\em J. Am. Chem. Soc.} \textbf{139}, 18262-18270 (2017).
\bibitem{JACS_140_9882_(2018)} J. Nishida, J. P. Breen, K. P. Lindquist, D. Umeyama, H. I. Karunadasa, \& M. D. Fayer. Dynamically disordered lattice in a layered Pb-I-SCN perovskite thin film probed by two-dimensional infrared spectroscopy. {\em J. Am. Chem. Soc.} \textbf{140} 9882-9890 (2018). 

\bibitem {ARPC_67_65_2016} Herz, L. M. Charge-carrier dynamics in organic-inorganic metal halide perovskites. {\em Annu. Rev. Phys. Chem.} \textbf{67}, 65–89 (2016).
\bibitem{JACS_136_13818_2014} Oga, H., Saeki, A., Ogomi, Y., Hayase, S. \& Seki, S. Improved understanding of the electronic and energetic landscapes of perovskite solar cells: high local charge carrier mobility, reduced recombination, and extremely shallow traps. {\em J. Am. Chem. Soc.} \textbf{136}, 13818–13825 (2014).
\bibitem{EES_7_2269_2014} Wehrenfennig, C., Liu, M., Snaith, H. J., Johnston, M. B. \& Herz, L. M. Charge-carrier dynamics in vapour-deposited films of the organolead halide perovskite $\rm CH_{3}NH_{3}PbI_{3-x}Cl_{x}$. {\em Ener. Envir. Sci.} \textbf{7}, 2269–2275 (2014).
\bibitem {Science_353_1409_2016} Zhu, H. {\em et al.} Screening in crystalline liquids protects energetic carriers in hybrid perovskites. {\em Science} \textbf{353}, 1409-1413 (2016).
\bibitem {Sci_Adv_3_e1701217_2017} Miyata, K. {\em et. al.} Large Polarons in lead halide perovskites. {\em Sci. Adv.} \textbf{3}, e1701217 (2017). 
\bibitem {PNAS_114_7519_2017} Chen, T. {\em et al.} Origin of long lifetime of band-edge charge carriers in organic-inorganic lead iodide perovskites. {\em Proc. Natl. Am. Soc.} \textbf{114}, 7519-7524 (2017). 
\bibitem {Nat_Comm_9_1971_2018} Batignani, G. {\em et al.} Probing femtosecond lattice displacement upon photo-carrier generation in lead halide perovskite. {\em Nat. Comm.} \textbf{9}, 1971 (2018). 
\bibitem {Sci_Adv_3_e1602388_2017} Wu, X. {\em et al.} Light-induced picosecond rotational disordering of the inorganic sublattice in hybrid perovskites. {\em Sci. Adv.} \textbf{3}, e1602388 (2017). 
\bibitem {Ener_Envi_Sci_11_101_2018} Ambrosio, F. {\em et al.} Origin of low electron-hole recombination rate in metal halide perovskites. {\em Ener. Envi. Sci.} \textbf{11}, 101-105 (2018). 
\bibitem {Nano_Lett_15_7794_2015} Zheng, F. {\em et al.} Rashba spin-orbit coupling enhanced carrier lifetime in $\rm CH_{3}NH_{3}PbI_{3}$. 
\bibitem {Sci_Adv_3_e1700704_2017} Zhai, Y. {\em et al.} Giant Rashba splitting in 2D organic-inorganic halide perovskites measured by transient spectroscopies. {\em Sci. Adv.} \textbf{3}, e1700704 (2017). 
\bibitem {Nat_Mat_16_115_2017} Hutter, E. M. {\em et al.} Direct-indirect character of the bandgap in methylammonium lead iodide perovskite. {\em Nat. Mat.} \textbf{16}, 115-120 (2017). 
\bibitem {JPCL_6_693_2015} Liu, S. {\em et al.} Ferroelectric domain wall induced band gap reduction and charge separation in organometal halide perovskite. {\em J. Phys. Chem. Lett.} \textbf{6}, 693-699 (2015). 
\bibitem {JPCL_6_3663_2015} Bakulin, A. A. {\em et al.} Real-time observation of organic cation reorientation in methylammonium lead iodide perovskites. {\em J. Phys. Chem. Lett.} \textbf{6}, 3663-3669 (2015). 
\bibitem {JACS_139_4068_2017} Selig, O. {\em et al.} Organic cation rotation and immobilization in pure and mixed methylammonium lead-halide perovskites. {\em J. Am. Chem. Soc.} \textbf{139}, 4068-4074 (2017). 
\bibitem {JPCL_9_895_2018} Taylor, V. C. A. {\em et al.} Investigating the role of the organic cation in formamidinium lead iodide perovskite using ultrafast spectroscopy. {\em J. Phys. Chem. Lett.} \textbf{9}, 895-901 (2018).
\bibitem {NANO_Lett_17_4151_2017} Park, M. {\em et al.} Critical role of methylammonium librational motion in methylammonium lead iodide ($\rm CH_{3}NH_{3}PbI_{3}$) perovskite photochemistry. {\em Nano. Lett.} \textbf{17}, 4151-4157 (2017).  
\bibitem {Nat_Comm_9_2525_2018} Park, M. {\em et al.} Excited-state vibrational dynamics toward the polaron in methylammonium lead iodide perovskite. {\em Nat. Comm.} \textbf{9}, 2525 (2018). 
\bibitem {Nat_Comm_8_687_2017} Kim, H. {\em et al.} Direct observation of mode-specific phonon-band gap coupling in methylammonium lead halide perovskites. {\em Nat. Comm.} \textbf{8}, 687 (2017).
\bibitem {Nat_Comm_8_376_2017} Richter, J. M., Branchi, F., Camargo, F. V. A., Zhao, B., Friend, R. H., Cerullo, G. \& Deschler, F. Ultrafast carrier thermalization in lead iodide perovskite probed with two-dimensional electronic spectroscopy. {\em Nat. Comm.} \textbf{8}, 376 (2017).
\bibitem {JPCL_8_3211_2017} Monahan, D. M., Guo, L., Lin, J., Dou, L., Yang, P. and Fleming, G. R. Room-temperature coherent optical phonon in 2D electronic spectra of  ($\rm CH_{3}NH_{3}PbI_{3}$) perovskite as a possible cooling bootleneck. {\em J. Phys. Chem. Lett.} \textbf{8}, 3211-3215 (2017). 
\bibitem {CPL_545_40_2012} Butkus. V, Zigmantas. D, Valkunas. L, \& Abramavicius. D Vibrational vs. electronic coherence in 2D spectrum of molecular systems. {\em Chem Phys Lett} \textbf{545}, 40–43 (2012).
\bibitem {JPCA_118_10259_2014} Egorova, D. Self-analysis of coherent oscillations in time-resolved optical signals. {\em J. Phys. Chem. A} \textbf{118}, 10259 (2014). 

\bibitem {Nat_Comm_10_482_2019} Guo, P. et al. Infrared-pump electronic-probe of methylammonium lead iodide reveals electronically decoupled organic and inorganic sublattices. {\em Nat. Comm.} \textbf{10}, 482 (2019).


\bibitem {Nano_Lett_16_3809_2016} Neukirch, A. J. {\em et al.} Polaron stabilization by cooperative lattice distortion and cation rotations in hybrid perovskite materials. {\em Nano. Lett.} \textbf{16}, 3809-3816 (2016). 
\bibitem {JCP_92_5397_1990} Beck, A. D. \& Edgecombe, K. E. A simple measure of electron localization in atomic and molecular systems. {\em J. Chem. Phys.} \textbf{92}, 5397 (1990).
\bibitem {JCP_81_6026_1984} Stevens, W. J., Basch, H. \& Krauss, M. Compact effective potentials and efficient shared exponent basis sets for the first and second row atoms. {\em J. Chem. Phys.} \textbf{81}, 6026 (1984). 
\bibitem {JPCA_108_11056_2004} Praprotnik, M., Janezi, D. \& Mavri, J. Temperature dependence of water vibrational spectrum: a molecular dynamics simulation study. {\em J. Phys. Chem. A} \textbf{108}, 11056 (2004). 
\bibitem {PCCP_15_6608_2013} Thomas, M., Brehm, M., Fligg, R., V\"ohringer, P. \& Kirchner, B. Computing vibrational spectra from ab initio molecular dynamics. {\em Phys. Chem. Chem. Phys.} \textbf{15}, 6608 (2013).
\bibitem {JPCB_119_8080_2015} Mishra, P., Vendrell, O. \& Santra, R. Ultrafast energy transfer from solvent to solute induced by subpicosecond highly intense thz pulses. {\em J. Phys. Chem. B} \textbf{119}, 8080–8086 (2015). 
\bibitem {JCP_120_1_2004} Strachan, A. Normal modes and frequencies from covariances in molecular dynamics or monte carlo simulations. {\em J. Chem. Phys.} \textbf{120}, 1–4 (2004). 
\bibitem {JCP_138_094311_2013} El-Amine Madjet, M., Li, Z., \& Vendrell, O. Ultrafast hydrogen migration in acetylene cation driven by non-adiabatic effects. {\em J. Chem. Phys.} \textbf{138}, 094311 (2013).  
\bibitem {JCC_14_1347_1993} Schmidt, W. M. {\em et al.} General Atomic and Molecular Electronic Structure System. {\em J. Compt. Chem.} \textbf{14}, 1347-1363 (1993). 
\bibitem {Science_350_1501_2015} Ishikawa, T. {\em et al.} Direct observation of collective modes coupled to molecular orbital-driven charge transfer. {\em Science} \textbf{350}, 1501 (2015).
\bibitem {SPP_162_432_(2015)} Prokhorenko, V. I. {\em et al.} Broadband electronic two-dimensional spectroscopy in the deep UV. {\em Springer Proc. Phys.} \textbf{162}, 432-435 (2015). 
\bibitem {Chem_Soc_Jap_45_1030_2016} Zhang. Y, et al. Preferential facet growth of methylammonium lead halide single crystalsPromoted by halide coordination. {\em Chem. Soc. Jap.} \textbf{45}, 1030-1032 (2016).
\bibitem {Ener_Envi_Sci_10_145_2017} Noel. N. K, et al. A low viscosity, low boiling point, clean solvent system for the rapid crystallisation of highly specular perovskite films. {\em Ener. Envi. Sci.} \textbf{10},145-152 (2017). 
 
  
\end{thebibliography}
%
\begin{addendum}
 \item This work was supported by the Max Planck Society and the Excellence Cluster ``The Hamburg Center for Ultrafast Imaging - Advanced Imaging of Matter" of the Deutsche Forschungsgemeinschaft.  The authors thank V. I. Prokhorenko for help with the 2D setup and providing 2D data analysis software, and Marcelo Carignano for helpful discussions on the excited state MD simulation. PKN and HJS T acknowledge the support from the UK Engineering and Physical Sciences Research Council (grant no EP/P032591/1).

\item[Competing Interests] H.J.S. is the co-founder and CSO of Oxford PV Ltd, a company that is commercializing perovskite photovoltaic technologies. 

\item[Correspondence] Correspondence and requests for theoretical part should be addressed to  Z.L.(email: zheng.li@mpsd.mpg.de) and M.E.M. (email: mmadjet@hbku.edu.qa), correspondence and requests for this project should be addressed to R.J.D.M (email: dwayne.miller@mpsd.mpg.de). 
\end{addendum} 
%

\end{document}